\documentstyle[amssymb,floats,aps,prb,epsf]{revtex}
\epsfclipon 
\begin{document}

\draft
\title{Gauge invariant dressed holon and spinon in doped cuprates}
\author{Shiping Feng, Tianxing Ma, and Jihong Qin}
\address{Department of Physics and Key Laboratory of Beam
Technology and Material Modification, Beijing Normal University,
Beijing 100875, China}
\date{Received 16 February 2003}

\maketitle

\begin{abstract}
We develop a partial charge-spin separation fermion-spin theory
implemented the gauge invariant dressed holon and spinon. In this
novel approach, the physical electron is decoupled as the gauge
invariant dressed holon and spinon, with the dressed holon
behaviors like a spinful fermion, and represents the charge degree
of freedom together with the phase part of the spin degree of
freedom, while the dressed spinon is a hard-core boson, and
represents the amplitude part of the spin degree of freedom, then
the electron single occupancy local constraint is satisfied.
Within this approach, the charge transport and spin response of
the underdoped cuprates is studied. It is shown that the charge
transport is mainly governed by the scattering from the dressed
holons due to the dressed spinon fluctuation, while the scattering
from the dressed spinons due to the dressed holon fluctuation
dominates the spin response.
\end{abstract}
\pacs{74.25.Ha, 74.25.Fy, 74.72.-h}

\bigskip
Very soon after the discovery of the high-temperature
superconductivity (HTSC) in doped cuprates, Anderson proposed a
scenario of HTSC based on the charge-spin separation in the
two-dimension (2D) \cite{pwa1}, where the internal degrees of
freedom of the electron are decoupled as the charge and spin
degrees of freedom, while the elementary excitations are not
quasi-particles but collective modes for the charge and spin
degrees of freedom, {\it i.e}., the holon and spinon, then these
holon and spinon might be responsible for the experimentally
observed nonconventional behavior of doped cuprates. Many unusual
features of doped cuprates are extensively studied following this
line within the 2D $t$-$J$ type model.

The decoupling of the charge and spin degrees of freedom of
electron is undoubtedly correct in the one-dimensional (1D)
interacting electron systems \cite{haldane}. In particular, the
typical behavior of the non-Fermi-liquid, {\it i.e.}, the absence
of the quasi-particle propagation and charge-spin separation, has
been demonstrated theoretically within the 1D $t$-$J$ model
\cite{maekawa}. Moreover, the holon and spinon as the real
elementary excitations in 1D cuprates has been observed directly
by the angle-resolved photoemission spectroscopy (ARPES)
experiment \cite{shen2}. However, the case in 2D is very complex
since there are many competing degrees of freedom \cite{kbse1}. As
a consequence, both experimental investigation and theoretical
understanding are extremely difficult. Among the unusual features
of doped cuprates, a hallmark is the charge transport
\cite{kbse1,ando1}, where the conductivity shows a non-Drude
behavior at low energies, and is carried by $x$ holes, with $x$ is
the hole doping concentration, while the resistivity exhibits a
linear temperature behavior over a wide range of temperatures. This
is strong experimental evidence supporting the notion of the
charge-spin separation, since not even conventional
electron-electron scattering would show the striking linear rise of
scattering rate above the Debye frequency, and if there is no the
charge-spin separation, the phonons should affect these properties
\cite{pwa4}. Furthermore, a compelling evidence for the charge-spin
separation in doped cuprates has been found from the experimental
test of the Wiedemann-Franz law, where a clear departure from the
universal Wiedemann-Franz law for the typical Fermi-liquid behavior
is observed \cite{hill}. In this case, a formal theory implemented
the gauge invariant holon and spinon, {\it i.e.}, the issue of
whether the holon and spinon are real, is centrally important, since
we must squarely face if there is to be a meaningful discussion of
theories based on the charge-spin separation \cite{laughlin}. In
this paper, we propose a partial charge-spin separation
fermion-spin theory, and show that if the local single occupancy
constraint is treated properly, then the electron operator can be
decoupled by introducing the dressed holon and spinon. These
dressed holon and spinon are gauge invariant, {\it i.e.}, they are
real in 2D.

We begin with the $t$-$t'$-$J$ model defined on a square lattice
as,
\begin{eqnarray}
H=-t\sum_{i\hat{\eta}\sigma}C^{\dagger}_{i\sigma}
C_{i+\hat{\eta}\sigma}+t'\sum_{i\hat{\tau}\sigma}
C^{\dagger}_{i\sigma}C_{i+\hat{\tau}\sigma}+\mu \sum_{i\sigma}
C^{\dagger}_{i\sigma}C_{i\sigma}+J\sum_{i\hat{\eta}}{\bf S}_{i}
\cdot {\bf S}_{i+\hat{\eta}},
\end{eqnarray}
supplemented by the single occupancy local constraint
$\sum_{\sigma}C^{\dagger}_{i\sigma}C_{i\sigma}\leq 1$,  where
$\hat{\eta}=\pm\hat{x},\pm\hat{y}$,
$\hat{\tau}=\pm\hat{x}\pm\hat{y}$, $C^{\dagger}_{i\sigma}$
($C_{i\sigma}$) is the electron creation (annihilation) operator,
${\bf S}_{i}=C^{\dagger}_{i}{\vec\sigma}C_{i}/2$ is spin operator
with ${\vec\sigma}=(\sigma_{x},\sigma_{y},\sigma_{z})$ as Pauli
matrices, and $\mu$ is the chemical potential. It has been shown
that the local constraint can be treated properly in analytical
calculations within the fermion-spin theory \cite{feng1}. In this
approach \cite{feng1}, the constrained electron operator is
decoupled in the CP$^{1}$ representation as
$C_{i\sigma}=h^{\dagger}_{i}a_{i\sigma}$, with the
local constraint $\sum_{\sigma}a^{\dagger}_{i\sigma}a_{i\sigma}=1$,
where the fermion operator $h_{i}$ keeps track of the charge degree
of freedom, while the boson operator $a_{i\sigma}$ keeps track of
the spin degree of freedom, then the electron local constraint
$\sum_{\sigma}C^{\dagger}_{i\sigma}C_{i\sigma}=1-h^{\dagger}_{i}
h_{i}\leq 1$ is satisfied, with $n_{h}=h^{\dagger}_{i}h_{i}$ is the
holon number at site $i$, equal to 1 or 0. In this formalism, the
charge and spin degrees of freedom of electron may be separated at
the mean field (MF) level, where the elementary charge and spin
excitations are called the holon and spinon, respectively. We call
such holon and spinon as bare holon and spinon, respectively, since
an extra $U(1)$ gauge degree of freedom related with the local
constraint $\sum_{\sigma}a^{\dagger}_{i\sigma}a_{i\sigma}=1$
appears, {\it i.e.}, the CP$^{1}$ representation is invariant under
a local $U(1)$ gauge transformation
$h_{i}\rightarrow h_{i}e^{i\theta_{i}}$,
$a_{i\sigma}\rightarrow a_{i\sigma}e^{i\theta_{i}}$, and then all
physical quantities should be invariant with respect to this
transformation. Thus both bare holon $h_{i}$ and bare spinon
$a_{i\sigma}$ are not gauge invariant, and they are strongly
coupled by this $U(1)$ gauge field fluctuations. In other words,
these bare holon and spinon are not real.

However, the CP$^{1}$ boson $a_{i\sigma}$ with the constraint can
be mapped exactly onto the pseudospin representation defined with
an additional phase factor. This is because that the empty and
doubly occupied spin states have been ruled out due to the
constraint $a^{\dagger}_{i\uparrow}a_{i\uparrow}+
a^{\dagger}_{i\downarrow}a_{i\downarrow}=1$, and only the spin-up
and spin-down singly occupied spin states are allowed, thus the
original four-dimensional representation space is reduced to a 2D
space. Due to the symmetry of the spin-up and spin-down states,
$\mid {\rm occupied} \rangle_{\uparrow}=\left (\matrix{1\cr
0\cr}\right)_{\uparrow}$ and $\mid {\rm empty} \rangle_{\uparrow}
=\left (\matrix{0\cr 1\cr}\right)_{\uparrow}$ are singly-occupied
and empty spin-up, while $\mid {\rm occupied}\rangle_{\downarrow}=
\left (\matrix{0\cr 1\cr}\right)_{\downarrow}$ and $\mid{\rm empty}
\rangle_{\downarrow}=\left (\matrix{1\cr 0\cr}\right)_{\downarrow}$
are singly-occupied and empty spin-down states, respectively. In
this case, the constrained CP$^{1}$ boson operators $a_{i\sigma}$
can be represented in this reduced 2D space as,
\begin{mathletters}
\begin{eqnarray}
a_{\uparrow}=e^{i\Phi_{\uparrow}}\mid {\rm occupied}
\rangle_{\downarrow}~_{\uparrow}\langle {\rm occupied}\mid=
e^{i\Phi_{\uparrow}}\left (\matrix{0 &0\cr 1 &0\cr}\right)
=e^{i\Phi_{\uparrow}}S^{-}, \\
a_{\downarrow}=e^{i\Phi_{\downarrow}}\mid {\rm occupied}
\rangle_{\uparrow}~_{\downarrow}\langle {\rm occupied}\mid=
e^{i\Phi_{\downarrow}}\left (\matrix{0 &1\cr 0 &0\cr}\right)
=e^{i\Phi_{\downarrow}}S^{+},
\end{eqnarray}
\end{mathletters}
with $S^{-}$ is the $S^{z}$ lowering operator, while $S^{+}$ is
the $S^{z}$ raising operator, and then the local constraint
$\sum_{\sigma}a^{\dagger}_{i\sigma}a_{i\sigma}=S^{+}_{i}S^{-}_{i}
+S^{-}_{i}S^{+}_{i}=1$ is exactly satisfied. Obviously, the bare
spinon contains both phase and amplitude parts, and the phase part
is described by the phase factor $e^{i\Phi_{i\sigma}}$, while the
amplitude part is described by the spin operator $S_{i}$. In this
case, the electron CP$^{1}$ decoupling form with the constraint
can be expressed as
$C_{i\uparrow}=h^{\dagger}_{i}e^{i\Phi_{i\uparrow}}S^{-}_{i}$ and
$C_{i\downarrow}=h^{\dagger}_{i}e^{i\Phi_{i\downarrow}}S^{+}_{i}$,
while the local $U(1)$ gauge transformation is replaced as
$h_{i}\rightarrow h_{i}e^{i\theta_{i}}$,
$\Phi_{i\sigma}\rightarrow \Phi_{i\sigma}+\theta_{i}$. Furthermore,
the phase factor of the bare spinon $e^{i\Phi_{i\sigma}}$ can be
incorporated into the bare holon, then we obtain a new fermion-spin
transformation as,
\begin{eqnarray}
C_{i\uparrow}=h^{\dagger}_{i\uparrow}S^{-}_{i},~~~~
C_{i\downarrow}=h^{\dagger}_{i\downarrow}S^{+}_{i},
\end{eqnarray}
with the spinful fermion operator
$h_{i\sigma}=e^{-i\Phi_{i\sigma}} h_{i}$ describes the charge
degree of freedom together with the phase part of the spin degree
of freedom (dressed holon), while the spin operator $S$ describes
the amplitude part of the spin degree of freedom (dressed spinon).
In this case, only the amplitude part of the spin degree of
freedom is separated from the electron operator, and thus the
partial charge-spin separation is implemented in the electron
decoupling form (3). These dressed holon and spinon are invariant
under the local $U(1)$ gauge transformation, and therefore all
physical quantities from the dressed holon or dressed spinon also
are invariant with respect to this gauge transformation. In this
sense, these dressed holon and spinon are real as the new
elementary particle excitations in the low-dimensional solid. This
gauge invariant dressed holon is a {\it magnetic dressing}. In
other words, the gauge invariant dressed holon carries some spinon
messages, {\it i.e.}, it shares some effects of the spinon
configuration rearrangements due to the presence of the hole
itself. We emphasize that the present dressed holon $h_{i\sigma}$
is a spinless fermion $h_{i}$ (bare holon) incorporated a spinon
cloud (magnetic flux), although in common sense $h_{i\sigma}$ is
not an real spinful fermion, its behaviors like a spinful fermion.
In correspondence with these special physical properties, we find
that $h^{\dagger}_{i\sigma}h_{i\sigma}=h^{\dagger}_{i}
e^{i\Phi_{i\sigma}}e^{-i\Phi_{i\sigma}}h_{i}=h^{\dagger}_{i}h_{i}$,
which guarantees that the electron on-site local constraint,
$\sum_{\sigma}C^{\dagger}_{i\sigma}C_{i\sigma}=S^{+}_{i}
h_{i\uparrow}h^{\dagger}_{i\uparrow}S^{-}_{i}+S^{-}_{i}
h_{i\downarrow}h^{\dagger}_{i\downarrow}S^{+}_{i}=h_{i}
h^{\dagger}_{i}(S^{+}_{i}S^{-}_{i}+S^{-}_{i}S^{+}_{i})=1-
h^{\dagger}_{i}h_{i}\leq 1$, is always satisfied in analytical
calculations. Moreover the double {\it spinful fermion} occupancy,
$h^{\dagger}_{i\sigma}h^{\dagger}_{i-\sigma}=e^{i\Phi_{i\sigma}}
h^{\dagger}_{i}h^{\dagger}_{i}e^{i\Phi_{i-\sigma}}=0$,
$h_{i\sigma}h_{i-\sigma}=e^{-i\Phi_{i\sigma}}h_{i}h_{i}
e^{-i\Phi_{i-\sigma}}=0$, are ruled out automatically. Since the
spinless fermion $h_{i}$ and spin operators $S^{+}_{i}$ and
$S^{-}_{i}$ obey the anticommutation relation and Pauli spin
algebra, respectively, it then is easy to show that the spinful
fermion $h_{i\sigma}$ also obey the same anticommutation relation
as the spinless fermion $h_{i}$. In this partial charge-spin
separation fermion-spin representation, the $t$-$t'$-$J$ model
(1) can be rewritten as,
\begin{eqnarray}
H&=&-t\sum_{i\hat{\eta}}(h_{i\uparrow}S^{+}_{i}
h^{\dagger}_{i+\hat{\eta}\uparrow}S^{-}_{i+\hat{\eta}}+
h_{i\downarrow}S^{-}_{i}h^{\dagger}_{i+\hat{\eta}\downarrow}
S^{+}_{i+\hat{\eta}})+t'\sum_{i\hat{\tau}}(h_{i\uparrow}S^{+}_{i}
h^{\dagger}_{i+\hat{\tau}\uparrow}S^{-}_{i+\hat{\tau}}+
h_{i\downarrow}S^{-}_{i}h^{\dagger}_{i+\hat{\tau}\downarrow}
S^{+}_{i+\hat{\tau}})\nonumber \\
&-&\mu\sum_{i\sigma}h^{\dagger}_{i\sigma}h_{i\sigma}
+{1\over 2}J\sum_{i\hat{\eta}\sigma}(h_{i\sigma}
h^{\dagger}_{i\sigma}){\bf S}_{i}\cdot{\bf S}_{i+\hat{\eta}}
(h_{i+\hat{\eta}-\sigma}h^{\dagger}_{i+\hat{\eta}-\sigma}).
\end{eqnarray}
The spirit of the present partial charge-spin separation
fermion-spin theory is very similar to these of the bosonization
in 1D interacting electron system \cite{haldane}, where the
electron operators are mapped onto the boson (electron density)
representation, and then the recast Hamiltonian is exactly
solvable.

Although the choice CP$^{1}$ representation is convenient, so long
as $h^{\dagger}_{i}h_{i}=1$,
$\sum_{\sigma}C^{\dagger}_{i\sigma}C_{i\sigma}=0$, no matter what
the values of $S^{+}_{i}S^{-}_{i}$ and $S^{-}_{i}S^{+}_{i}$ are,
therefore it means that a "spin" even to an empty site has been
assigned. It has been shown \cite{feng1} that this defect can be
cured by the projection operator $P_{i}$, {\it i.e.}, the
constrained electron operator can be mapped exactly using the
fermion-spin transformation defined with the additional projection
operator $P_{i}$ as $C_{i\uparrow}=P_{i}h^{\dagger}_{i\uparrow}
S^{-}_{i}P^{\dagger}_{i}$ and $C_{i\downarrow}=P_{i}
h^{\dagger}_{i\downarrow}S^{+}_{i}P^{\dagger}_{i}$. However, this
projection operator is cumbersome to handle, and we will drop it in
the actual calculations. It also has been shown \cite{feng1,feng2}
that such treatment leads to errors of the order $x$ in counting
the number of spin states, which is negligible for small dopings.
Moreover, the electron local constraint still is exactly obeyed
even in the MF approximation (MFA), and therefore the essential
physics of the gauge invariant dressed holon and spinon also are
kept. This is because that the constrained electron operator
$C_{i\sigma}$ in the $t$-$J$ type model also can be mapped onto the
slave-fermion formulism \cite{feng2} as
$C_{i\sigma}=h^{\dagger}_{i}b_{i\sigma}$ with the local constraint
$h^{\dagger}_{i}h_{i}+\sum_{\sigma}b^{\dagger}_{i\sigma}b_{i\sigma}
=1$. We can solve this constraint by rewriting the boson operators
$b_{i\sigma}$ in terms of the CP$^{1}$ boson operators
$a_{i\sigma}$ as
$b_{i\sigma}=a_{i\sigma}\sqrt{1-h^{\dagger}_{i}h_{i}}$ supplemented
by the local constraint
$\sum_{\sigma}a^{\dagger}_{i\sigma}a_{i\sigma}=1$. As mentioned
above, the CP$^{1}$ boson operators $a_{i\uparrow}$ and
$a_{i\downarrow}$ with the local constraint can be identified with
the pseudospin lowering and raising operators, respectively,
defined with an additional phase factor,  therefore the projection
operator is approximately related to the holon number operator by
$P_{i}\sim\sqrt{1-h^{\dagger}_{i\sigma}h_{i\sigma}}=
\sqrt{1-h^{\dagger}_{i}h_{i}}$, and its main role is to remove the
spurious spin when there is a holon at the site $i$.

In the Fermi-liquid, the electron carried both charge and spin
degrees of freedom moves almost freely since the weak
electron-electron interaction in the system, where both amplitude
and phase parts of the spin degree of freedom are delocalized.
However, in the doped Mott insulator, our present study indicates
that the electron operator can be decoupled as the dressed holon
and spinon, with the dressed spinon represents the bare spinon's
amplitude part, and is localized, while the dressed holon
represents the bare holon together with the bare spinon's phase
part, and then can move almost freely in the background of the
dressed spinon fluctuation. The present decoupling scheme seems to
have been confirmed in the doped 1D Mott insulator, where although
the charge and spin degrees of freedom are decoupled, the charge
and spin degrees of freedom are manifested themself by the
excitations of charge-density wave and spin-density wave,
respectively \cite{haldane}. These charge-density wave and
spin-density wave are described in terms of the density-density
correlation function and spin-spin correlation, respectively, and
both density operator and spin operator are gauge invariant.

As an application of the present theory, we discuss the charge
transport and spin response of the underdoped cuprates. The
one-particle dressed holon and spinon two-time Green's functions
are defined as,
\begin{mathletters}
\begin{eqnarray}
g_{\sigma}(i-j,t-t')&=&-i\theta(t-t')\langle [h_{i\sigma}(t),
h^{\dagger}_{j\sigma}(t')]\rangle=\langle\langle h_{i\sigma}(t);
h^{\dagger}_{j\sigma}(t')\rangle\rangle ,\\
D(i-j,t-t')&=&-i\theta(t-t')\langle [S^{+}_{i}(t),S^{-}_{j}(t')]
\rangle =\langle\langle S^{+}_{i}(t);S^{-}_{j}(t')\rangle\rangle,
\end{eqnarray}
\end{mathletters}
respectively. Since the dressed spinon operators obey the Pauli
algebra, then our goal is to evaluate the dressed holon and spinon
Green's functions directly for the fermion and spin operators in
terms of equation of motion method. In the framework of equation of
motion, the time-Fourier transform of the two-time Green's function
$G(\omega)=\langle\langle A;A^{\dagger}\rangle\rangle_{\omega}$
satisfies the equation \cite{zubarev}, $\omega\langle\langle
A;A^{\dagger}\rangle\rangle_{\omega}=\langle [A,A^{\dagger}]\rangle
+\langle\langle [A,H];A^{\dagger}\rangle\rangle_{\omega}$. If we
define the orthogonal operator $L$ as $[A,H]=\zeta A-iL$, with
$\langle [L,A^{\dagger}]\rangle =0$, then the full Green's function
can be expressed as,
\begin{eqnarray}
G(\omega)=G^{(0)}(\omega)+{1\over \varsigma^{2}}G^{(0)}(\omega)
\langle\langle L;L^{\dagger}\rangle\rangle_{\omega}G^{(0)}(\omega),
\end{eqnarray}
with the MF Green's function
$G^{(0)}(\omega)=\varsigma/(\omega-\zeta)$, where
$\varsigma=\langle [A,A^{\dagger}]\rangle$. It has been shown
\cite{zubarev} that if the self-energy $\Sigma(\omega)$ is
identified as the irreducible part of
$\langle\langle L;L^{\dagger}\rangle\rangle_{\omega}$, then the
full Green's function (5) can be evaluated as,
\begin{eqnarray}
G(\omega)={\varsigma\over \omega-\zeta-\Sigma(\omega)},
\end{eqnarray}
with $\Sigma(\omega)=\langle\langle L;L^{\dagger}\rangle
\rangle^{irr}_{\omega}/\varsigma$. In the framework of the
diagrammatic technique, $\Sigma(\omega)$ corresponds to the
contribution of irreducible diagrams.

It has been shown from the experiments \cite{kbse6} that the
antiferromagnetic (AF) long-range-order (AFLRO) in the undoped
cuprates is destroyed by hole doping of the order $\sim 0.024$,
therefore there is no AFLRO in the underdoped regime $0.025\leq x<
0.15$, {\it i.e.}, $\langle S^{z}_{i}\rangle =0$. In this case, a
MF theory \cite{feng3} of the $t$-$J$ model has been discussed
within the Kondo-Yamaji decoupling scheme \cite{kondo}. Following
their discussions \cite{feng3}, we can obtain the MF dressed holon
and spinon Green's functions in the present case as,
$g^{(0)}_{\sigma}({\bf k},\omega)=1/(\omega-\xi_{k})$ and
$D^{(0)}({\bf k},\omega)=B_{k}/(\omega^{2}-\omega_{k}^{2})$,
respectively, where $B_{k}=\lambda_{1}[2\chi^{z}_{1}(\epsilon
\gamma_{{\bf k}}-1)+\chi_{1}(\gamma_{{\bf k}}-\epsilon)]-
\lambda_{2}(2\chi^{z}_{2}\gamma'_{{\bf k}}-\chi_{2})$,
$\lambda_{1}=2ZJ_{eff}$, $\lambda_{2}=4Z\phi_{2}t'$, $\phi_{1}=
\langle h^{\dagger}_{i\sigma}h_{i+\hat{\eta}\sigma}\rangle$,
$\phi_{2}=\langle h^{\dagger}_{i\sigma}h_{i+\hat{\tau}\sigma}
\rangle$, $\epsilon=1+2t\phi_{1}/J_{{\rm eff}}$,
$J_{{\rm eff}}=(1-x)^{2}J$, $x=\langle h^{\dagger}_{i\sigma}
h_{i\sigma}\rangle=\langle h^{\dagger}_{i}h_{i}\rangle$,
$\gamma_{{\bf k}}=(1/Z)\sum_{\hat{\eta}}e^{i{\bf k}\cdot
\hat{\eta}}$, $\gamma'_{{\bf k}}=(1/Z)\sum_{\hat{\tau}} e^{i{\bf
k}\cdot\hat{\tau}}$, $Z$ is the number of the nearest neighbor or
second-nearest neighbor sites, the MF dressed holon spectrum,
$\xi_{k}=Zt\chi_{1}\gamma_{{\bf k}}-Zt'\chi_{2} \gamma'_{{\bf
k}}-\mu$, and the MF dressed spinon spectrum,
$\omega^{2}_{k}=A_{1}(\gamma_{k})^{2}+A_{2}(\gamma'_{k})^{2}+
A_{3}\gamma_{k}\gamma'_{k}+A_{4}\gamma_{k}+A_{5}\gamma'_{k}+A_{6}$,
with $A_{1}=\alpha\epsilon\lambda_{1}^{2}(\epsilon\chi^{z}_{1}+
\chi_{1}/2)$, $A_{2}=\alpha\lambda_{2}^{2}\chi^{z}_{2}$, $A_{3}=-
\alpha\lambda_{1}\lambda_{2}(\epsilon\chi^{z}_{1}+\epsilon
\chi^{z}_{2}+\chi_{1}/2)$, $A_{4}=-\epsilon\lambda_{1}^{2}[\alpha
(\chi^{z}_{1}+\epsilon\chi_{1}/2)+(\alpha C^{z}_{1}+(1-\alpha)/(4Z)
-\alpha\epsilon\chi_{1}/(2Z))+(\alpha C_{1}+(1-\alpha)/(2Z)-\alpha
\chi^{z}_{1}/2)/2]+\alpha\lambda_{1}\lambda_{2}(C_{3}+\epsilon
\chi_{2})/2$, $A_{5}=-3\alpha\lambda^{2}_{2}\chi_{2}/(2Z)+\alpha
\lambda_{1}\lambda_{2}(\chi^{z}_{1}+\epsilon\chi_{1}/2+C^{z}_{3})$,
$A_{6}=\lambda^{2}_{1}[\alpha C^{z}_{1}+(1-\alpha)/(4Z)-\alpha
\epsilon\chi_{1}/(2Z)+\epsilon^{2}(\alpha C_{1}+(1-\alpha)/(2Z)-
\alpha\chi^{z}_{1}/2)/2]+\lambda^{2}_{2}(\alpha C_{2}+(1-\alpha)/
(2Z)-\alpha\chi^{z}_{2}/2)/2)-\alpha\epsilon\lambda_{1}\lambda_{2}
C_{3}$, and the spinon correlation functions $\chi_{1}=\langle
S_{i}^{+}S_{i+\hat{\eta}}^{-}\rangle$, $\chi^{z}_{1}=\langle
S_{i}^{z}S_{i+\hat{\eta}}^{z}\rangle$, $\chi_{2}=\langle
S_{i}^{+}S_{i+\hat{\tau}}^{-}\rangle$, $\chi^{z}_{2}=\langle
S_{i}^{z}S_{i+\hat{\tau}}^{z}\rangle$,
$C_{1}=(1/Z^{2})\sum_{\hat{\eta},\hat{\eta'}}\langle
S_{i+\hat{\eta}}^{+}S_{i+\hat{\eta'}}^{-}\rangle$,
$C^{z}_{1}=(1/Z^{2})\sum_{\hat{\eta},\hat{\eta'}}\langle
S_{i+\hat{\eta}}^{z}S_{i+\hat{\eta'}}^{z}\rangle$,
$C_{2}=(1/Z^{2})\sum_{\hat{\tau},\hat{\tau'}}\langle
S_{i+\hat{\tau}}^{+}S_{i+\hat{\tau'}}^{-}\rangle$,
$C_{3}=(1/Z)\sum_{\hat{\tau}}\langle S_{i+\hat{\eta}}^{+}
S_{i+\hat{\tau}}^{-}\rangle$,  and  $C^{z}_{3}=(1/Z)
\sum_{\hat{\tau}}\langle S_{i+\hat{\eta}}^{z}
S_{i+\hat{\tau}}^{z}\rangle$. In order not to violate the sum rule
of the correlation function $\langle
S^{+}_{i}S^{-}_{i}\rangle=1/2$ in the case without AFLRO, the
important decoupling parameter $\alpha$ has been introduced in the
MF self-consistent calculation \cite{feng3,kondo}, which can be
regarded as the vertex correction.

With the help of Eq. (7), the full dressed holon and spinon
Green's functions are obtained as,
\begin{mathletters}
\begin{eqnarray}
g_{\sigma}({\bf k},\omega)&=&{1\over \omega-\xi_{k}-
\Sigma^{(2)}_{h}({\bf k},\omega)}, \\
D({\bf k},\omega)&=&{B_{k}\over \omega^{2} -\omega^{2}_{k}-
\Sigma^{(2)}_{s}({\bf k},\omega)},
\end{eqnarray}
\end{mathletters}
respectively, where the second-order dressed holon self-energy
from the dressed spinon pair bubble
$\Sigma^{(2)}_{h}({\bf k},\omega)=\langle\langle L^{(h)}_{k}(t);
L^{(h)\dagger}_{k}(t')\rangle\rangle_{\omega}$ with the orthogonal
operator $L^{(h)}_{i}=-t\sum_{\hat{\eta}}h_{i+\hat{\eta}\sigma}
(S^{-}_{i+\hat{\eta}}S^{+}_{i}-\chi_{1})+t'\sum_{\hat{\tau}}
h_{i+\hat{\tau}\sigma}(S^{-}_{i+\hat{\tau}}S^{+}_{i}-\chi_{2})$,
and can be evaluated as,
\begin{eqnarray}
\Sigma_{h}^{(2)}({\bf k},\omega)&=&{1\over 2}\left ({Z\over N}
\right )^2\sum_{pp'}\gamma^{2}_{12}({\bf k,p,p'}){B_{p'}B_{p+p'}
\over 4\omega_{p'}\omega_{p+p'}}\left ({F^{(h)}_{1}(k,p,p')\over
\omega+\omega_{p+p'}-\omega_{p'}-\xi_{p+k}}+{F^{(h)}_{2}(k,p,p')
\over\omega+\omega_{p'}-\omega_{p+p'}-\xi_{p+k}}\right.\nonumber \\
&+&\left.{F^{(h)}_{3}(k,p,p')\over\omega+\omega_{p'}+\omega_{p+p'}
-\xi_{p+k}}-{F^{(h)}_{4}(k,p,p')\over\omega -\omega_{p+p'}-
\omega_{p'}-\xi_{p+k}}\right ),
\end{eqnarray}
where $\gamma^{2}_{12}({\bf k,p,p'})=[(t\gamma_{{\bf p'+p+k}}-t'
\gamma'_{{\bf p'+p+k}})^{2}+(t\gamma_{{\bf p'-k}}-t'
\gamma'_{{\bf p'-k}})^{2}]$, $F^{(h)}_{1}(k,p,p')=n_{F}(\xi_{p+k})
[n_{B}(\omega_{p'})-n_{B}(\omega_{p+p'})]+n_{B}(\omega_{p+p'})[1+
n_{B}(\omega_{p'})]$, $F^{(h)}_{2}(k,p,p')=n_{F}(\xi_{p+k})[n_{B}
(\omega_{p'+p})-n_{B}(\omega_{p'})]+n_{B}(\omega_{p'})[1+n_{B}(
\omega_{p'+p})]$, $F^{(h)}_{3}(k,p,p')=n_{F}(\xi_{p+k})[1+n_{B}
(\omega_{p+p'})+n_{B}(\omega_{p'})]+n_{B}(\omega_{p'})n_{B}
(\omega_{p+p'})$, $F^{(h)}_{4}(k,p,p')=n_{F}(\xi_{p+k)}[1+n_{B}
(\omega_{p+p'})+n_{B}(\omega_{p'})]-[1+n_{B}(\omega_{p'})][1+n_{B}
(\omega_{p+p'})]$, and $n_{B}(\omega_{p})$ and $n_{F}(\xi_{p})$ are
the boson and fermion distribution functions, respectively. The
calculation of the dressed spinon self-energy is quite tedious,
since our start point is the dressed spinon MF solution within the
Kondo-Yamaji decoupling scheme. The full dressed spinon Green's
function satisfies the relation,
$\omega^{2}D(k,\omega)=B_{k}+\langle\langle
[[S^{+}_{i}(t),H(t)],H(t)];S^{-}_{j}(t')\rangle\rangle_{k,\omega}$
with $[[S^{+}_{i},H],H]_{k}=\omega^{2}_{k}S^{+}_{k}-iL^{(s)'}_{k}$.
In the disordered spin liquid state without AFLRO, the dressed
holon-spinon interaction should dominate the essential physics. In
this case, the orthogonal operator for the dressed spinon can be
selected as \cite{feng6},
\begin{eqnarray}
L^{(s)}_{i}&=&-(2\epsilon\chi^{z}_{1}+\chi_{1})\lambda_{1}
{1\over Z}\sum_{\hat{\eta},\hat{a}}t_{\hat{a}}
(h^{\dagger}_{i+\hat{\eta}\uparrow}h_{i+\hat{\eta}+\hat{a}\uparrow}
+h^{\dagger}_{i+\hat{\eta}+\hat{a}\downarrow}
h_{i+\hat{\eta}\downarrow}-2\phi_{\hat{a}})
S^{+}_{i+\hat{\eta}+\hat{a}}+[(2\chi^{z}_{1}+\epsilon\chi_{1})
\lambda_{1}-\chi_{2}\lambda_{2}]\sum_{\hat{a}}t_{\hat{a}}
(h^{\dagger}_{i\uparrow}h_{i+\hat{a}\uparrow}\nonumber \\
&+&h^{\dagger}_{i+\hat{a}\downarrow}h_{i\downarrow}-
2\phi_{\hat{a}})S^{+}_{i+\hat{a}}+2\chi^{z}_{2}\lambda_{2}
{1\over Z}\sum_{\hat{\tau},\hat{a}}t_{\hat{a}}
(h^{\dagger}_{i+\hat{\tau}\uparrow}h_{i+\hat{\tau}+\hat{a}\uparrow}
+h^{\dagger}_{i+\hat{\tau}+\hat{a}\downarrow}
h_{i+\hat{\tau}\downarrow}-2\phi_{\hat{a}})
S^{+}_{i+\hat{\tau}+\hat{a}},
\end{eqnarray}
where $\hat{a}=\hat{\eta},\hat{\tau}$, with $t_{\hat{\eta}}=t$,
$\phi_{\hat{\eta}}=\phi_{1}$, and $t_{\hat{\tau}}=-t'$,
$\phi_{\hat{\tau}}=\phi_{2}$. After a straightforward calculation,
we obtain the dressed spinon self-energy $\Sigma^{(2)}_{s}({\bf k},
\omega)=\langle\langle L^{(s)}_{i}(t);L^{(s)\dagger}_{j}(t')\rangle
\rangle_{k,\omega}$ as,
\begin{eqnarray}
\Sigma_{s}^{(2)}({\bf k},\omega)=B_{k}\left ({Z\over N}
\right )^{2}\sum_{pp'}\gamma^{2}_{12}({\bf k,p,p'})
{B_{k+p}\over 2\omega_{k+p}}\left ({F^{(s)}_{1}(k,p,p')\over\omega+
\xi_{p+p'}-\xi_{p'}-\omega_{k+p}}-{F^{(s)}_{2}(k,p,p')\over \omega+
\xi_{p+p'}-\xi_{p'}+\omega_{k+p}}\right ),
\end{eqnarray}
with $F^{(s)}_{1}(k,p,p')=n_{F}(\xi_{p+p'})[1-n_{F}(\xi_{p'})]-
n_{B}(\omega_{k+p})[n_{F}(\xi_{p'})-n_{F}(\xi_{p+p'})]$, and
$F^{(s)}_{2}(k,p,p')=n_{F}(\xi_{p+p'})[1-n_{F}(\xi_{p'})]+[1+n_{B}
(\omega_{k+p})][n_{F}(\xi_{p'})-n_{F}(\xi_{p+p'})]$. Within the
diagrammatic technique, this dressed spinon self-energy
$\Sigma_{s}^{(2)}({\bf k},\omega)$ corresponds to the contribution
from the dressed holon pair bubble.

In the present partial charge-spin separation theoretical
framework, the external electronic field can only be coupled to the
gauge invariant dressed holons, but the strong correlation between
dressed holons and spinons still is considered self-consistently
through the dressed spinon's order parameters entering in the
dressed holon's propagator. In this case, the resistivity can be
obtained as $\rho=1/\sigma_{dc}$, with $\sigma_{dc}=
\lim_{\omega\rightarrow 0}\sigma(\omega)$, and the optical
conductivity \cite{feng6},
\begin{eqnarray}
\sigma(\omega)=({Ze\over 2})^2{1\over N}\sum_{k\sigma}
\gamma_{s}^{2}({\bf k})\int^{\infty}_{-\infty}{d\omega'\over 2\pi}
A^{(h)}_{\sigma}({\bf k},\omega'+\omega)A^{(h)}_{\sigma}({\bf k},
\omega'){n_{F}(\omega'+\omega)-n_{F}(\omega') \over \omega},
\end{eqnarray}
where $\gamma^{2}_{s}({\bf k})=[\sin^{2} k_{x}(\chi_{1}t-2\chi_{2}
t'\cos k_{y})^{2}+\sin^{2} k_{y}(\chi_{1}t-2\chi_{2}
t'\cos k_{x})^{2}]/4$ and the dressed holon spectral function
$A^{(h)}_{\sigma}({\bf k},\omega)=-2{\rm Im}g_{\sigma}({\bf k},
\omega)$. The results of the resistivity as a function of
temperature at $x =0.03$ (solid line), $x=0.04$ (dashed line),
$x=0.05$ (dotted line), and $x=0.06$ (dash-dotted line) for
$t/J=2.5$ and $t'/t=0.15$ are plotted in Fig. 1 in comparison with
the experimental data \cite{ando1} taken from
La$_{2-x}$Sr$_{x}$CuO$_{4}$ (inset). It is shown obviously that
the resistivity is characterized by a crossover from the moderate
temperature metallic-like behavior to low temperature
insulating-like behavior in the heavily underdoped regime, and a
temperature linear dependence with deviations at low temperatures
in the moderate underdoped regime. But even in the heavily
underdoped regime, the resistivity exhibits the metallic-like
behavior over a wide range of temperatures, which are in agreement
with the experiments \cite{ando1}.
\begin{figure}[ht]
\epsfxsize=4.5in\centerline{\epsffile{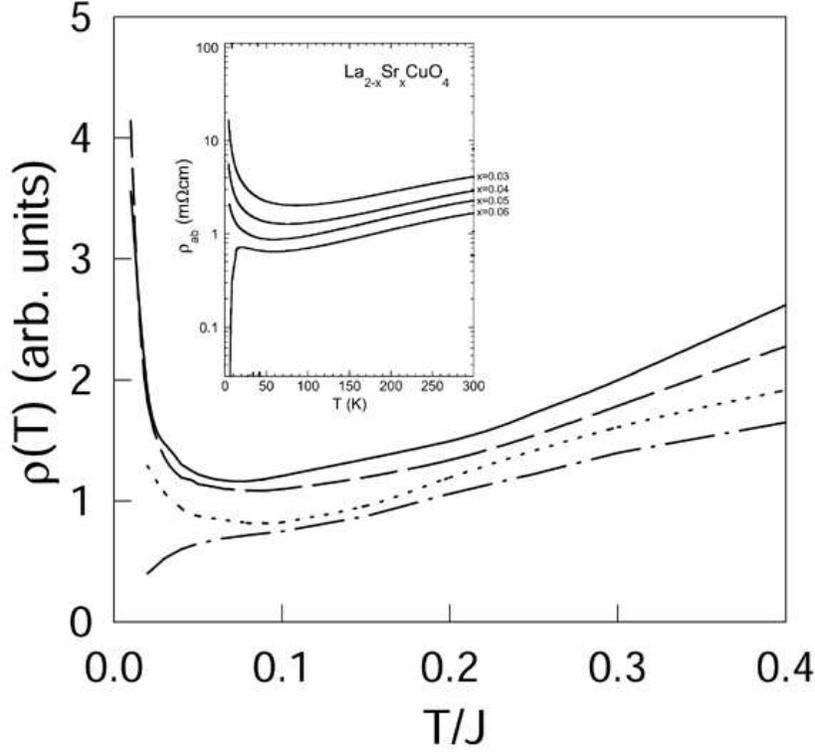}}
\caption{The electron resistivity as a function of temperature
at $x=0.03$ (solid line), $x=0.04$ (dashed line), $x=0.05$
(dotted line), and $x=0.06$ (dash-dotted line) with $t/J=2.5$
and $t'/t=0.15$. Inset: the experimental result of
La$_{2-x}$Sr$_{x}$CuO$_{4}$ taken from Ref. \cite{ando1}.}
\end{figure}

The doped holes into the Mott insulator can be considered as a
competition between the kinetic energy ($xt$) and magnetic energy
($J$). The magnetic energy $J$ favors the magnetic order for spins
and results in "frustration" of the kinetic energy, while the
kinetic energy $xt$ favors delocalization of holes and tends to
destroy the magnetic order. In the present partial charge-spin
separation fermion-spin theory, the scattering of dressed holons
dominates the charge transport. The dressed holon scattering rate
is obtained from the dressed holon self-energy by considering the
dressed holon-spinon interaction, while this dressed holon
self-energy from the dressed spinon pair bubble characterizes the
competition between the kinetic energy and magnetic energy,
therefore the striking behavior in the resistivity is intriguingly
related with this competition. In the heavily underdoped regime,
the dressed holon kinetic energy is much smaller than the dressed
spinon magnetic energy in lower temperatures, therefore the
dressed holons are localized, and scattering rate from the dressed
holon self-energy is severely reduced, this leads to the
insulating-like behavior in the resistivity. With increasing
temperatures, the dressed holon kinetic energy is increased, while
the dressed spinon magnetic energy is decreased. In the region
where the dressed holon kinetic energy is larger than the dressed
spinon magnetic energy at moderate temperatures, the dressed
holons move almost freely, and then the dressed holon scattering
would give rise to the metallic-like behavior in the resistivity.

Now we discuss the dynamical spin response. In the present partial
charge-spin separation fermion-spin theory, the spin fluctuation
couples only to dressed spinons, but the effect of dressed holons
on dressed spinons has been considered through the dressed holon's
order parameters entering in the dressed spinon propagator. In
this case, we can obtain the dynamical spin structure factor
$S({\bf k},\omega)=-2[1+n_{B}(\omega)]{\rm Im}D({\bf k},\omega)$
as,
\begin{eqnarray}
S({\bf k},\omega)=-2[1+n_{B}(\omega)]{B_{k}{\rm Im}\Sigma_{s}^{(2)}
({\bf k},\omega)\over [\omega^{2}-\omega^{2}_{k}-{\rm Re}
\Sigma_{s}^{(2)}({\bf k},\omega)]^{2}+[{\rm Im}\Sigma_{s}^{(2)}
({\bf k},\omega)]^{2}},
\end{eqnarray}
with ${\rm Im}\Sigma_{s}^{(2)}({\bf k},\omega)$ and
${\rm Re}\Sigma_{s}^{(2)}({\bf k},\omega)$ are corresponding
imaginary part and real part of the dressed spinon self-energy
function $\Sigma_{s}^{(2)}({\bf k},\omega)$.

We plot the dynamical spin structure factor spectrum
$S({\bf k},\omega)$ in the ($k_{x},k_{y}$) plane at $x=0.06$ with
$T=0.05J$ and $\omega=0.05J$ for $t/J=2.5$ and $t'/t=0.15$ in
Fig. 2.
\begin{figure}[ht]
\epsfxsize=4.5in\centerline{\epsffile{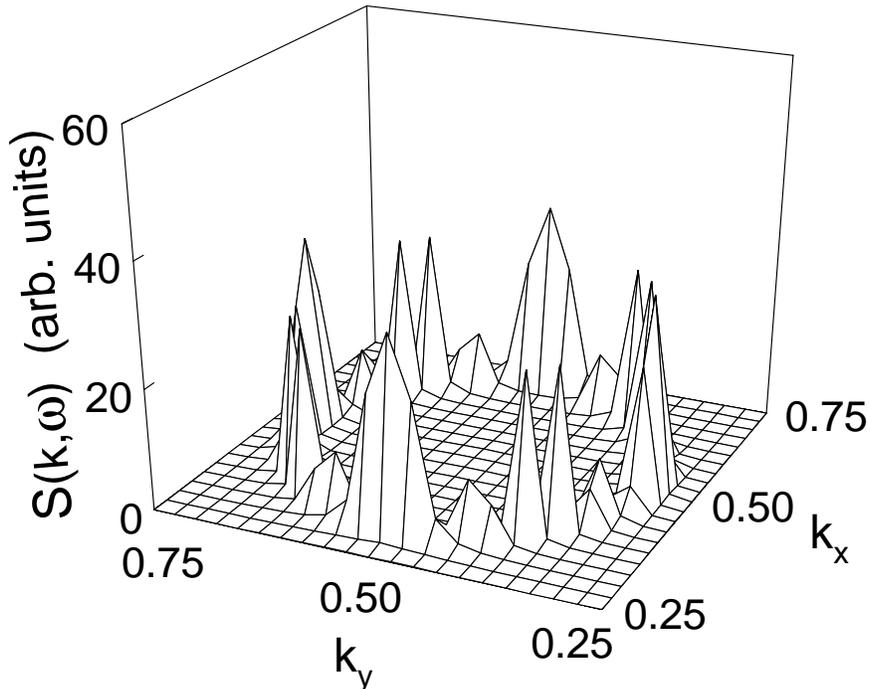}}
\caption{The dynamical spin structure factor spectrum in the
$(k_{x},k_{y})$ plane at $x=0.06$ in $T=0.05J$ and
$\omega=0.05J$ with $t/J=2.5$ and $t'/t=0.15$.}
\end{figure}
It is shown that with dopings, there is a
commensurate-incommensurate (IC) transition in the spin fluctuation
geometry, where all IC peaks lie on a circle of radius of $\delta$.
Although some IC satellite diagonal peaks appear, the main weight
of the IC peaks is in the parallel direction, and these parallel
peaks are located at $[(1\pm\delta)/2,1/2]$ and
$[1/2,(1\pm \delta)/2]$ (hereafter we use the units of
$[2\pi,2\pi]$). The present dynamical spin structure factor
spectrum $S({\bf k},\omega)$ has been used to extract the doping
dependence of the incommensurability $\delta(x)$, which is defined
as the deviation of the peak position from AF wave vector position,
and the result is shown in Fig. 3 in comparison with the
experimental result \cite{kbse3} taken from
La$_{2-x}$Sr$_{x}$CuO$_{4}$ (inset). Our results show that
$\delta(x)$ increases progressively with the hole concentration at
lower dopings, but saturates at higher dopings, which are
qualitatively consistent with the experimental observations
\cite{kbse3}.
\begin{figure}[ht]
\epsfxsize=4.5in\centerline{\epsffile{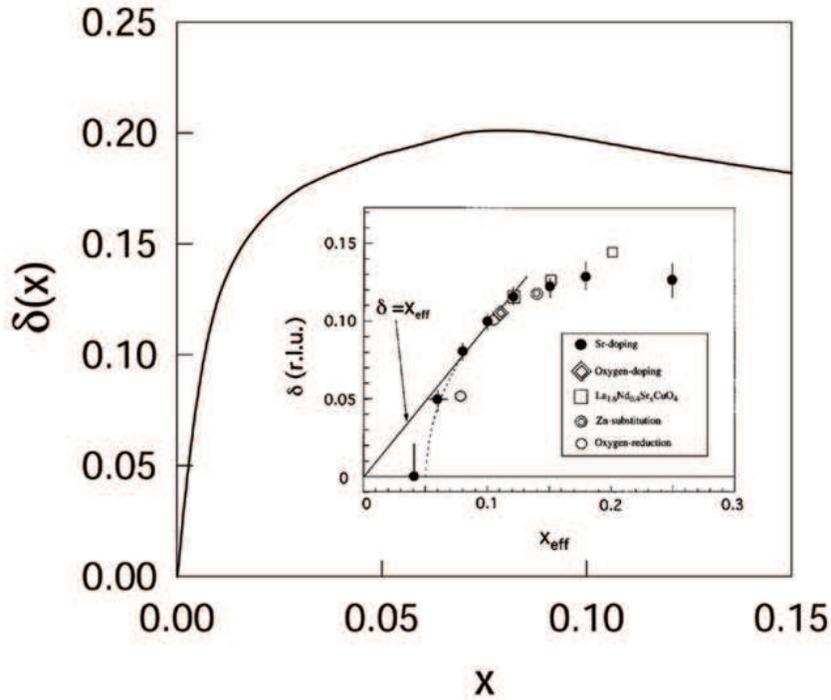}}
\caption{The doping dependence of the incommensurability
$\delta(x)$. Inset: the experimental result of
La$_{2-x}$Sr$_{x}$CuO$_{4}$ taken from Ref. \cite{kbse3}.}
\end{figure}

Our results also show that the effect of dressed holons on the
dressed spinon part is critical in determining the characteristic
feature of the IC magnetic correlation. This can be understood from
the properties of the dressed spinon excitation spectrum
$E^{2}_{k}=\omega^{2}_{k}+{\rm Re}\Sigma^{(2)}_{s}({\bf k},E_{k})$.
During the calculation of the dynamical spin structure factor
spectrum $S({\bf k},\omega)$, we find when
$W({\bf k}_{\delta},\omega)=[\omega^{2}-\omega^{2}_{k_{\delta}}-
{\rm Re}\Sigma_{s}^{(2)}({\bf k}_{\delta},\omega)]^{2}\sim 0$ at
the critical wave vectors $\pm{\bf k}_{\delta}$ in low energies,
the IC peaks appear, then the weight of the IC peaks is dominated
by the inverse of the imaginary part of the dressed spinon
self-energy $1/{\rm Im}\Sigma_{s}^{(2)}({\bf k}_{\delta},\omega)$.
In this case, the positions of the IC peaks are determined by both
functions $W({\bf k},\omega)$ and ${\rm Im}\Sigma_{s}^{(2)}({\bf
k},\omega)$, where the zero points of $W({\bf k},\omega)$ (then
the critical wave vectors ${\bf k}_{\delta}$) is doping
dependence. Near the half-filling, the zero point of $W({\bf
k},\omega)$ locates at the AF wave vector [$1/2,1/2$], so the
commensurate AF peak appears there. With doping, the holes disturb
the AF background. As a result of self-consistent motion of dressed
holons and spinons, the IC magnetic correlation is developed beyond
certain critical doping. As seen from $S({\bf k},\omega)$, the
physics is dominated by the dressed spinon self-energy
renormalization due to dressed holon pair bubble. In this sense,
the mobile dressed holons are the key factor leading to the IC
magnetic correlation, {\it i.e.}, the mechanism of the IC type of
structure in doped cuprates away from the half-filling is most
likely related to the dressed holon motion. This is why the
position of the IC peaks can be determined in the present study
within the $t$-$t'$-$J$ model, while the dressed spinon energy
dependence is ascribed purely to the self-energy effects which
arise from the dressed holon-spinon interaction.

In summary, we have developed a partial charge-spin separation
fermion-spin theory to study the physical property of doped
cuprates. In this novel approach, the electron operator is
decoupled as the dressed holon and spinon, with the dressed holon
keeps track of the charge degree of freedom together with the
phase part of the spin degree of freedom, while the dressed
spinon keeps track of the amplitude part of the spin degree of
freedom, then the electron single occupancy local constraint is
satisfied even in MFA. The dressed holon is a magnetic dressing,
and then its behaviors like a spinful fermion, while the dressed
spinon is neither boson nor fermion, but a hard-core boson.
Moreover, both dressed holon and spinon are gauge invariant, and
in this sense, they are real as the new elementary particle
excitations in the low-dimensional solid. Within the $t$-$t'$-$J$
model, we have studied the charge transport and spin response
of the underdoped cuprates, and results are qualitatively
consistent with the experimental observations. These results also
are qualitatively consistent with the previous results
\cite{feng4} based on the fermion-spin theory \cite{feng1}, where
the phase factor $e^{i\Phi_{i\sigma}}$ described the phase part of
the spin degree of freedom was not considered. Our results also
show that the charge transport is mainly governed by the
scattering from the dressed holons due to the dressed spinon
fluctuation, while the scattering from the dressed spinons due to
the dressed holon fluctuation dominates the spin response. In this
case, the charge transport and spin response are almost
independent, and the perturbations that interact primarily with
charge do not much affect spin, therefore the notion of the
partial charge-spin separation naturally accounts for the
qualitative features of doped cuprates.

The present partial charge-spin separation fermion-spin theory
also indicates that the kinetic energy ($t$) term in the $t$-$J$
type model gives the dressed holon and spinon dynamics in the
doped regime without AFLRO, while the magnetic energy ($J$) term
is only to form an adequate dressed spinon configuration. Although
the projection operator has been dropped in the actual
calculations, leading to an overcounting the number of spin states,
the calculated results show that it does not matter in the
low-energy sector \cite{pwa4}, this is because that the effect of
the kinetic energy on the spinon configuration has been considered
self-consistently in the partial charge-spin separation
fermion-spin theory.

Irrespective of the coupling mechanism responsible for HTSC in
doped cuprates, the superconducting state is characterized by the
electron Cooper pairs \cite{tsuei}. It has been shown \cite{shen3}
from ARPES that in the real space the gap function and pairing
force have a range of one lattice spacing. In this case, the
superconducting order parameter for the electron pair can be
expressed in the present theory as $\Delta_{\hat{\eta}}=\langle
C^{\dagger}_{i\uparrow}C^{\dagger}_{i+\hat{\eta}\downarrow}-
C^{\dagger}_{i\downarrow}C^{\dagger}_{i+\hat{\eta}\uparrow}\rangle
=\langle h_{i\uparrow}h_{i+\hat{\eta}\downarrow}S^{+}_{i}
S^{-}_{i+\hat{\eta}}-h_{i\downarrow}h_{i+\hat{\eta}\uparrow}
S^{-}_{i}S^{+}_{i+\hat{\eta}}\rangle$. In the doped regime without
AFLRO, the dressed spinons form the disordered liquid state. In
this case, $\langle S^{+}_{i}S^{-}_{i+\hat{\eta}}\rangle=\langle
S^{-}_{i}S^{+}_{i+\hat{\eta}}\rangle$, and $\Delta_{\hat{\eta}}=
\Delta^{(h)}_{\hat{\eta}}\langle S^{+}_{i}S^{-}_{i+\hat{\eta}}
\rangle$ with the dressed holon Cooper pair
$\Delta^{(h)}_{\hat{\eta}}=\langle h_{i\uparrow}
h_{i+\hat{\eta}\downarrow}-h_{i\downarrow}h_{i+\hat{\eta}\uparrow}
\rangle$, which shows that the symmetry of the electron Cooper pair
is essentially determined by the symmetry of the dressed holon
Cooper pair \cite{feng5}. It then is possible that the kinetic
energy terms in the higher powers of the doping concentration $x$
cause the superconductivity \cite{pwa4}. Since this form of the
electron Cooper pair is common, then there is a coexistence of the
electron Cooper pair and the AF short-range fluctuation. In other
words, the AF short-range correlation can persist into the
superconducting state, which is in agreement with the experiments
\cite{kbse1}. This also shows that the AF short-range correlation
may play an important role in the mechanism for HTSC
\cite{kbse1,pwa1,pwa4}. However, in the doped regime with AFLRO,
where $\langle S^{+}_{i}S^{-}_{i+\hat{\eta}}\rangle\neq\langle
S^{-}_{i}S^{+}_{i+\hat{\eta}}\rangle$, then obviously the magnetic
correlation with AFLRO is not favorable for the superconductivity.

\acknowledgments
The authors would like to thank Professor Z.Q. Huang, Professor
Z.B. Su, Professor L. Yu, Dr. F. Yuan, and Professor Z.X. Zhao for
the helpful discussions. This work was supported by the National
Natural Science Foundation of China under Grant Nos. 10125415 and
10074007.


\begin{references}

\bibitem {pwa1} P.W. Anderson, Science {\bf 235}, 1196 (1987).

\bibitem {haldane} F.D.M. Haldane, Phys. Rev. Lett. {\bf 45}, 1358
(1980); Phys. Lett. {\bf 81}A, 153 (1981).

\bibitem {maekawa} See, {\it e.g.}, S. Maekawa and T. Tohyama,
Rep. Prog. Phys. {\bf 64}, 383 (2001).

\bibitem {shen2} C. Kim {\it et al.}, Phys. Rev. Lett. {\bf 77},
4054 (1996).

\bibitem{kbse1} See, {\it e.g.}, M.A.Kastner {\it et al.}, Mod.
Phys. {\bf 70}, 897 (1998), and referenes therein.

\bibitem{ando1} Y. Ando {\it et al.}, Phys. Rev. Lett. {\bf 87},
017001 (2001).

\bibitem {pwa4} P.W. Anderson, Science {\bf 288}, 480 (2000);
Phys. Rev. Lett. {\bf 67}, 2092 (1991).

\bibitem {hill} R.W. Hill {\it et al.}, Nature {\bf 414}, 711
(2001).

\bibitem {laughlin} R.B. Laughlin, Phys. Rev. Lett. {\bf 79},
1726 (1997).

\bibitem {feng1} Shiping Feng, Z.B. Su, and L. Yu, Phys. Rev. B
{\bf 49}, 2368 (1994); Mod. Phys. Lett. B{\bf 7}, 1013 (1993).

\bibitem {feng2} Shiping Feng and Feng Yuan, in {\it Symposium on
the Frontiers of Physics at Millennium}, edited by Y.L. Wu and
J.P. Hsu (World Scientific, Singapore, 2001), p. 221.

\bibitem {zubarev} D.N. Zubarev, Sov. Phys. Usp. {\bf 3}, 201
(1960).

\bibitem {kbse6} B. Keimer {\it et al.}, Phys. Rev. B {\bf 46},
14034 (1992); M. Matsuda {\it et al.}, Phys. Rev. B {\bf 62},
9148 (2000).

\bibitem {feng3} Shiping Feng and Yun Song, Phys. Rev. B {\bf 55},
642 (1997).

\bibitem{kondo} J. Kondo and K. Yamaji, Prog. Theor. Phys.
{\bf 47}, 807 (1972).

\bibitem {feng6} Shiping Feng, Jihong Qin, and Tianxing Ma
(unpublished).

\bibitem {kbse3} K. Yamada {\it et al.}, Phys. Rev. B {\bf 57},
6165 (1998), and referenes therein.

\bibitem {feng4} Shiping Feng and Zhongbing Huang, Phys. Lett.
A{\bf 232}, 293 (1997); Shiping Feng and Zhongbing Huang, Phys.
Rev. B {\bf 57}, 10328 (1998); Feng Yuan {\it et al.}, Phys. Rev.
B {\bf 64}, 224505 (2001).

\bibitem{tsuei} See, e.g., C.C. Tsuei and J.R. Kirtley, Rev. Mod.
Phys. {\bf 72}, 969 (2000), and referenes therein.

\bibitem{shen3} Z.X. Shen {\it et al.}, Phys. Rev. Lett. {\bf 70},
1553 (1993); H. Ding {\it et al.}, Phys. Rev. B{\bf 54}, R9678
(1996).

\bibitem{feng5} Shiping Feng, Phys. Lett. A{\bf 257}, 325 (1999).

\end{references}
\end{document}